\documentclass[aps,prl,twocolumn]{revtex4}
\usepackage{amssymb}

\usepackage{graphicx}

\begin{document}

\title{Detecting the breached pair phase in a polarized ultracold Fermi gas}
\author{W. Yi and L.-M. Duan}
\address{FOCUS center and MCTP, Department of Physics, University of Michigan, Ann
Arbor, MI 48109}

\begin{abstract}
We propose a method for the experimental detection of a new quantum
phase, the breached pair state, in a strongly interacting ultracold
Fermi gas with population imbalance. We show that through the
time-of-flight Raman imaging, the presence of such a phase can be
unambiguously determined with a measurement of the momentum-space
phase separation of the minority spin component. To guide the
experimental efforts, the momentum-space density profiles are
calculated under typical experimental conditions.
\end{abstract}

\maketitle

The search for new quantum phases has long been a persistent goal
of many-body physics. Ultracold atomic gas, with its remarkable
controllability, has provided a powerful platform to realize and
probe such phases. Among the various many-body phases, a few have
been realized in this system, including, for instance, the Bose or
the Fermi superfluid states \cite{1}, and the Mott insulator state
in an optical lattice \cite{2}. Here, we propose a method for the
experimental detection of a new quantum phase, the breached pair
state, in a strongly interacting gas of fermionic atoms with
population imbalance. Such a system, with its recent experimental
realization \cite {3,4,4a}, has raised strong interest \cite
{5,5a,6,6a,7,8,9,mc,10,10a,10b,10c}. In particular, a quantum
phase transition from a BCS superfluid to a normal state has been
observed \cite{3,4a}.

The breached pair (BP) state, proposed in Refs. \cite{5,5a,6} and
studied in detail in \cite{7,8,9,10}, is a non-BCS superfluid
phase with gapless fermionic excitations. A unique feature of this
state is its phase separation in the momentum space, although in
real space it is a homogeneous, polarized superfluid. This feature
also distinguishes the breached pair phase from a simple mixture
of molecular condensate and fermionic atoms, where there is no
phase separation in the momentum space. As illustrated in Fig. 1,
there are two kinds of BP states, with one or two Fermi surfaces
for the excess fermions \cite{5a,9}, respectively. To distinguish
the two scenarios, we simply label them the BP1 and the BP2
states. Except for the number of Fermi surface, the properties of
these two states are very similar. The quasi-particle excitations
in both of the BP phases have non-monotonic dispersion relations
\cite{5,5a,8}. There exists a well-defined quantum phase
transition from the BCS superfluid to either one of the two
phases. Such a phase transition is not characterized by symmetry
breaking, but rather by the change in the topology of the Fermi
surface \cite{9,10}.

\begin{figure}[tbp]
\includegraphics[height=5cm,width=8cm]{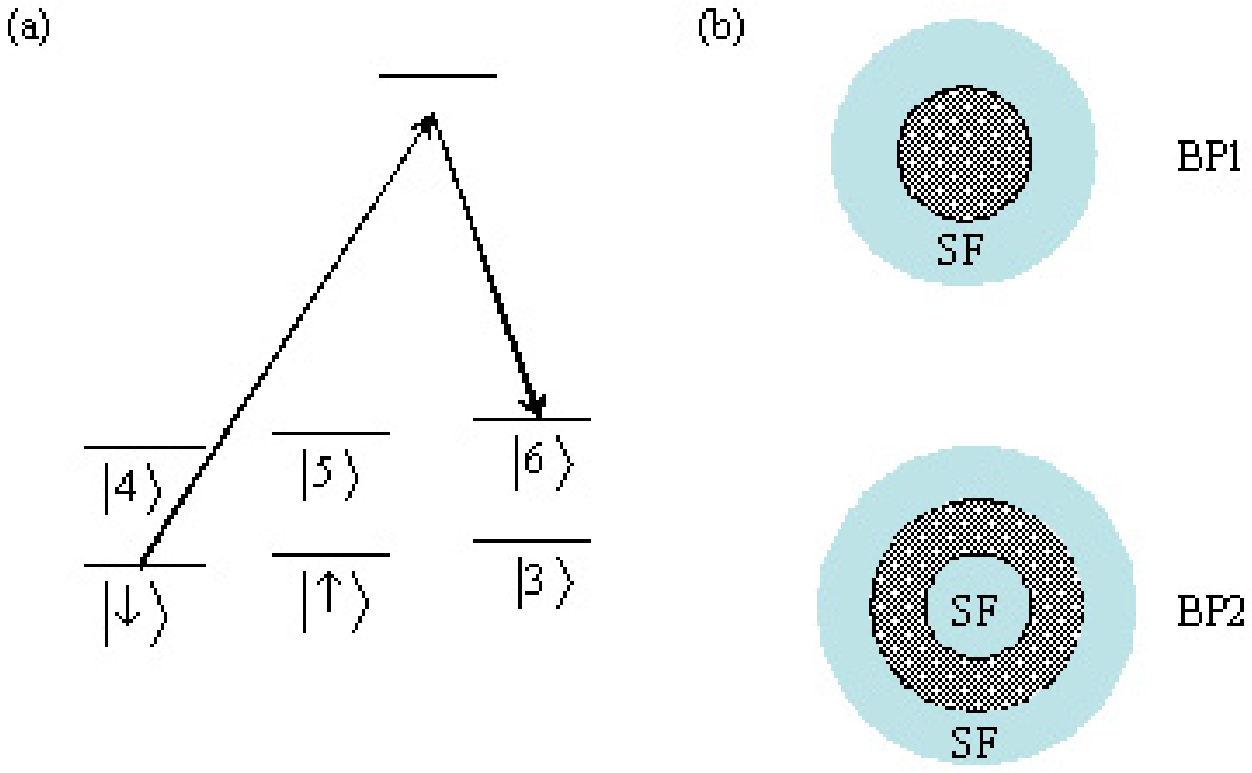}
\includegraphics[height=3.4cm,width=8cm]{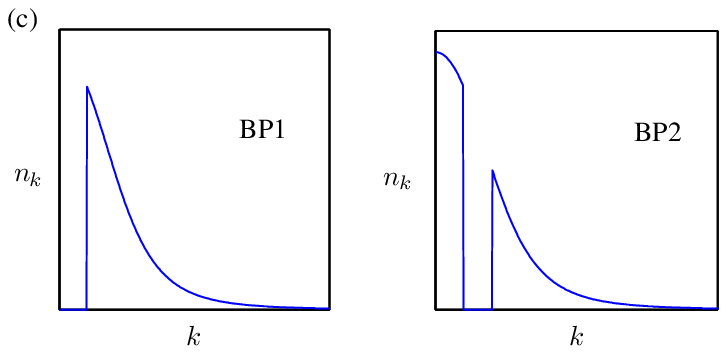}
\caption[Fig.1 ]{ (a) The level configuration for the $^{6}$Li
atoms near the Feshbach resonance point $B_{0}=834G$. An impulsive
$\protect\pi$ Raman pulse can selectively transfer the atoms from
the lowest level ($\left| \downarrow \right\rangle $) to the
highest level ($\left| 6\right\rangle $) in the ground-state
manifold. (b) Schematic momentum-space configurations of fermions
in the BP1 (top) and the BP2 (bottom) phases. The shaded regions
with dots are filled with the unpaired fermions of the majority
spin component, while the regions labelled 'SF' are filled with
the gapless superfluid. (c) Schematic momentum-space density
profiles of the minority fermions in the BP1 (left) and the BP2
(right) phase.}
\end{figure}

The BP phases have been predicted to exist in theory, but it is hard
to detect them experimentally. These phases show polarized
superfluidity. However, even if one can confirm polarized
superfluidity in the future experiments, it is not an unambiguous
signal for the BP phase. Many phases can have polarized
superfluidity. In particular, as we have shown in \cite{9}, the
conventional BCS state in the polarized gas also gives polarized
superfluidity at any finite temperature (which is necessarily the
case for experiments) due to the polarized quasi-particle
excitations. So, it is hard to distinguish the more exotic phases
from the conventional thermal BCS superfluid in experiments.

To overcome this difficulty, we propose a method to detect the BP
phases by measuring the momentum-space density profile of the
minority spin component. This method has the following desirable
features: first, it can give an unambiguous signal for the BP
phases. From the basic physical picture of the BP states shown in
Fig. 1(b), we see already that there exists a finite region in the
momentum-space with no minority atoms. This region, enclosed by
the Fermi surface(s), is occupied only by the majority spin
component. Therefore, if one measures the momentum-space density
profiles, the profile for the minority atoms should be
non-monotonic with a dip in the momentum distribution. This dip,
if observed, is an unambiguous signature of the BP phase as no
other phases have that character. Second, this method can also
unambiguously distinguish different kinds of BP states. In Fig.
1(c), we show schematically the characteristic momentum
distributions for the minority atoms associated with different BP
phases. The different structures of the dips in the momentum
distributions clearly tell the nature of the corresponding phases.
More detailed calculations of the momentum space profiles later in
the paper will further confirm the feasibility of this detection
method.

For non-interacting atomic gases, the time-of-flight imaging
provides a way to measure the momentum-space density profile. The
absorption image of the expanded cloud is proportional to the
initial atomic momentum distribution (after column integration).
However, to realize the BP states, the atoms need to be strongly
interacting in the trap. One therefore has to devise a method to
suddenly turn off this strong interaction right before the
expansion. For $^{40}$K atoms, such an interaction turn-off can be
achieved with a fast switch of the magnetic field, as has been
demonstrated in the recent experiment \cite{J10}. The experiments
on the polarized Fermi gases are done so far with $^{6}$Li atoms,
whose inter-atomic interaction, due to the large resonance width
of $^6$Li at $B_0=834G$, is hard to be turned off through the
magnetic filed switching without introducing significant influence
on the momentum distribution of the atoms.

To overcome this problem, we propose a method to tune the atomic
interaction almost instantaneously by applying an impulsive Raman
pulse consisting of two co-propagating laser beams. For $^{6}$Li
atoms close to a Feshbach resonance (the magnetic field ranging from
$600$ G to $1000$ G), the energy eigen-levels in the ground-state
manifold are shown in Fig. 1(a). The three lower (or upper) levels
differ mainly by their respective nuclear spins, while the states
between the two sets have different electronic spins. The Feshbach
resonance at $B_0=834G$ resonantly couples the lowest two energy
levels ($\left| \downarrow \right\rangle$ and $\left| \uparrow
\right\rangle $) in the ground state manyfold. Near this resonance
point, the intra-set level splitting (e.g., between the states
$\left| \downarrow \right\rangle $ and $\left| \uparrow
\right\rangle $) is about $70$ MHz, while the inter-set splitting
(e.g., between the states $\left| \downarrow \right\rangle $ and
$\left| 6\right\rangle $) is around $2.5$ GHz. We assume that the
atoms are initially in the levels $\left| \downarrow \right\rangle $
and $ \left| \uparrow \right\rangle $, with spin $\downarrow $ as
the minority component. Right before the time-of-flight imaging, we
apply an impulsive Raman pulse to fully transfer the minority
component from the state $\left| \downarrow \right\rangle $ to
$\left| 6\right\rangle $ (see Fig. 1(a)). As the level $\left|
6\right\rangle $ is detuned from $\left| \downarrow \right\rangle $
by $2.5$ GHz, which is larger than the width of the Feshbach
resonance at $B_0$ ($2\mu _{B}W/\hbar \sim 0.84$ GHz, where $W=300G$
is the width of the Feshbach resonance at $B_0=834G$), it does not
participate in the resonant coupling and thus only weakly interacts
with the spin $\left| \uparrow \right\rangle $ component. Since the
gas is dilute and the atoms in the two different states $\left|
\uparrow \right\rangle $ and $\left| 6\right\rangle $ can be quickly
separated, as a good approximation, the momentum distribution of the
$\left| 6\right\rangle $ component has a negligible variation during
the free expansion and can be recorded with the absorption image. As
the Raman pulse consists of co-propagating laser beams, the momentum
distribution of the atoms are not changed during the transfer.
Therefore, using the method outlined above, one effectively measures
the in-trap momentum-space density profile for the minority atoms.

We add a few remarks on this detection method before going on to
the theoretical calculation. Firstly, in this scheme the Raman
pulse is impulsive with its duration significantly smaller than
$\hbar /\mu $, where $\mu $ is the characteristic chemical
potential of the gas. The dynamics of the system during such a
short interval is negligible, and thus the Raman transfer
faithfully keeps track of the momentum distribution of the atoms.
As a result of the impulsiveness of the Raman pulse, this method
does not resolve the energy spread in the atomic gas, and is
therefore different from the Bragg or the r.f. spectroscopy
\cite{11,rf}. The two methods are actually complementary to each
other, as the spectroscopy measures the energy spread while the
Raman imaging detects the momentum distribution. Secondly, an
impulsive Raman pulse without resolving the energy spread may have
a practical advantage as it is insensitive to small variation of
the energy splitting between the levels $\left| \downarrow
\right\rangle $ and $\left| 6\right\rangle $ caused by the
fluctuation of the magnetic field. Finally, we can also let the
two laser beams of the Raman pulse propagate along different
directions (say, along $x$ and $-x$ axis). Such a pulse will not
change the momentum distribution in the $y-z$ plane, while
imprinting a momentum kick along the $x$ direction, which helps to
quickly separate the atomic components $\left| \uparrow
\right\rangle $ and $\left| 6\right\rangle $, and thus further
reduces the influence of interaction during the expansion.

To guide the experimental effort, in the following we include some
calculations on the momentum-space density profile of the minority
atoms. The calculations show that on the BEC side of the resonance
\cite{BECnote}, a non-monotonic momentum profile with a dip in the
center should be visible, even if we take average in real space
over all the phases in the trap \cite{note1} and perform a column
integration in the momentum space. Our calculation method \cite{9}
is a generalization of the self-consistent $G_{0}G$ diagram scheme
\cite{12} to the case with unequal spin population, which at zero
temperature reduces to the mean-field approach \cite{7,8,9,10}
(the latter recovers the crossover theory in the equal population
case \cite{12}). We use the local density approximation for a
three-dimensional (anisotropic) harmonic trap, and pick up the
phase corresponding to the global minimum of the thermodynamic
potential. The detailed formalism has been described in Refs.
\cite{9}. The properties of the system are universally
characterized by three dimensionless parameters, the population
imbalance $\beta \equiv \left( N_{\uparrow }-N_{\downarrow
}\right) /\left( N_{\uparrow }+N_{\downarrow }\right) $, the
dimensionless interaction strength $k_{F}a_{s}$, and the
temperature $T/T_{F}$ , where $k_{F}=\sqrt{2mE_{F}/\hbar ^{2}}$
and $T_{F}=E_{F}/k_{B}$ are defined through the Fermi energy
$E_{F}$ at the center of the trap for a non-interacting Fermi gas
with equal spin populations under the local density approximation.
>From this definition, $E_{F}=(3N\hbar ^{3}\omega _{x}\omega
_{y}\omega _{z})^{\frac{1}{3}}$, where $N\equiv N_{\uparrow
}+N_{\downarrow }$ is the total particle number, $\omega _{i}$
($i=x,y,z$) is the trapping frequency along the $i$th direction.
The atomic scattering length $a_s$ is connected with the magnetic
field detuning $\Delta B=B-B_{0}$ through $a_{s}=a_{bg}(1-W/\Delta
B)$, where $a_{bg}$ is the background scattering length and $W$ is
the width of the Feshbach resonance.

The phase diagram of this system has been shown in \cite{9}, according to
which the BP1 phase appears on the BEC side of the resonance with $%
(k_{F}a_{s})^{-1}\gtrsim 0.5$ at low temperature. To calculate the momentum
space density profile, however, we need to average over all the phases in
the trap \cite{note2}. Under the local density approximation, the atom
number distribution $n_{\mathbf{kr}}$ depends on both the atomic momentum $%
\mathbf{k}$ and the position in the trap $\mathbf{r}$. The
momentum-space density profile is given by $n_{\mathbf{k}}=\int
n_{\mathbf{kr}}d^{3}\mathbf{r}$. At $(k_{F}a_{s})^{-1}=2$, $\beta
=0.8$, and at zero temperature, this density profile
$n_{\mathbf{k}}$ is plotted in Fig. 2 for the minority spin
component (spin $\downarrow $), together with its column integration $n_{%
\mathbf{k}}^{\prime }=\int n_{\mathbf{k}}dk_{z}.$ It is clear that
a non-monotonic behavior should be visible for this density
profile, even after the column integration \cite{note3}. The dip
is smaller than the one in the schematic figure 1(c) as one needs
to average over different phases here in the trap. With the above
parameters, there is a large region of the BP1 phase near the
center of the trap (there is still a small BCS superfluid core
right at the center), which is surrounded by a
normal gas of the majority spin component. To achieve the interaction strength $%
(k_{F}a_{s})^{-1}=2$, one can take a moderately negative detuning
with $\Delta B=-80$ G for the $^{6}$Li atoms, where the
system still has a long lifetime. This parameter then corresponds to $%
N=1.5\times 10^{5}$ atoms in a trap with typical trap frequencies $\omega
_{x}=\omega _{y}\sim 110Hz$ and $\omega _{z}\sim 23Hz$ \cite{3} (or with $%
N=1.2\times 10^{6}$ atoms in a slightly weaker trap with the trap
frequencies $\omega _{i}$ reduced by a factor of $2$).

\begin{figure}[tbp]
\includegraphics{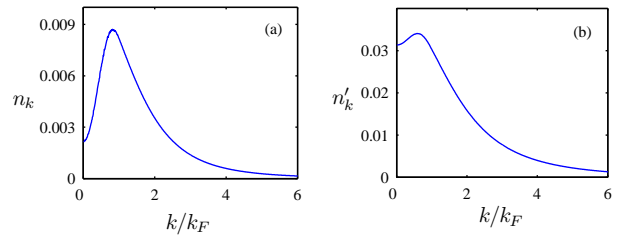}
\caption[Fig.2 ]{ (a) The momentum-space density profile $
n_{\mathbf{k}}$ (only dependent on $k=|\mathbf{k}|$) and (b) its
column integration $n_{\mathbf{k}}^{\prime }$ for the minority
spin component. The unit of the density is $n_{F}=k_{F}^{3}/(3%
\protect\pi ^{2})$, where $k_{F}$ has been specified in the text.
}
\end{figure}

\begin{figure}[tbp]
\includegraphics{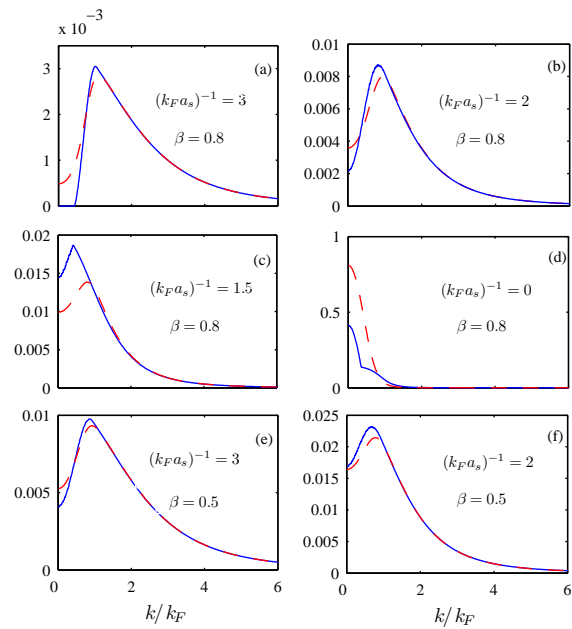}
\caption[Fig.3 ]{The momentum-space density profile $
n_{\mathbf{k}}$ at different temperatures, interaction strengths,
and population imbalance. The solid lines represent the
zero-temperature distributions, and the dashed ones represent the
distributions at $T=0.2T_{F}$. The interaction strength $
(k_{F}a_{s})$ and the population imbalance $\beta $ are specified
in the figure. Note that the hump at small momenta for the solid
line in panel (d) indicates the existence of a normal mixed phase
(NM) in the trap \cite{9}.}
\end{figure}

To check stability of the signal for the momentum-space phase
separation even at finite temperature, and to verify that this
signal is indeed uniquely connected with the BP phase, we
calculate the momentum-space density profiles $n_{\mathbf{k}}$ at
different interaction strengths ($ k_{F}a_{s}$) and temperatures
($T/T_{F}$), and with different population imbalance parameters
($\beta $). The results are shown in Fig. 3, from which several
observations can be made. First of all, this signal (the
non-monotonic profile $ n_{\mathbf{k}}$) remains stable at finite
but low temperatures. For instance, at $T=0.2T_{F}$ (which
corresponds a real temperature $T\sim 50nK$ for $N=1.5\times
10^{5}$ atoms in a trap with $\omega _{x}=\omega _{y}\sim 110Hz$
and $\omega _{z}\sim 23Hz$), the dip in the profile only becomes a
bit shallower. At significantly higher temperatures, the BP state
is eventually replaced by a normal phase, and the dip signal
disappears. Secondly, when one goes slightly further to the BEC
side \cite{note4}, such as with $ (k_{F}a_{s})^{-1}=3$, the dip
signal becomes more pronounced, which corresponds to a larger BP
phase region at the trap center. On resonance and on the BCS side,
the dip disappears (the dip first disappears on the BEC side where
the BP phase ceases to exist in the trap \cite{9}). Finally, as
one varies the interaction strength $k_{F}a_{s}$ or the population
imbalance $\beta $, the dip size varies, which correlates with the
variations in the size of the BP region in the phase diagram. This
shows that the dip signal is indeed uniquely connected with the BP
phase.

Before ending the paper, a few remarks are in order. Firstly, in
our calculation, only the BP1 phase can be stabilized in this
system, and it only appears on the BEC side of the resonance
\cite{note5}. Our proposed detection method, however, can detect
both the BP1 and the BP2 phases, and an experimental detection
surely can scan the whole region across the Feshbach resonance.
Similar to the real-space density profile \cite{3,4,4a}, a
momentum-space density profile will help us to understand a lot
about this strongly interacting system with various competing
phases, whether the experiment confirms the theoretical prediction
here or finds new surprise. Secondly, we use a single impulsive
Raman pulse here to fast tune the atomic interaction. If one
applies two consecutive impulsive Raman pulses propagating along
different directions, as shown in Ref. \cite{13}, one can
reconstruct the full real-space or momentum-space correlation
function through Fourier sampling with the laser phase gradient.
The full correlation function can reveal very detailed information
of the system, e.g., the tiny spatial structures in the system.
With such a capability, one may find some other interesting
quantum phases, such as the FFLO state \cite{14,14a,10b}, if the
latter does show up in this atomic gas.

We thank Cheng Chin and Martin Zwierlein for helpful discussions.
This work was supported by the NSF awards (0431476), the ARDA
under ARO contracts, and the A. P. Sloan Fellowship.


\begin{thebibliography}{99}
\bibitem{BECnote}  Note that the parameter region discussed in this
work is still close to the resonant point of a wide Feshbach
resonance, with $0<(k_Fa_s)^{-1}\le3$, where the fraction of the
close channel molecules is much less than $1\%$. So the wavefunction
of the bosonic molecules only has negligible influence on the
system.
\bibitem{note1}  Besides the BP phase, there are in general several
separated phases in a trap as one moves from the trap center to
the edge. See \cite{9} for the phase diagram.
\bibitem{note2}  We assume here that the Raman pulse transfers all the
minority atoms in the trap to the spectator state $\left| 6\right\rangle $.
If one can focus this laser beam to transfer atoms only in a particular
region (corresponding to the BP phase region) to the state $\left|
6\right\rangle $, one can have a more sensitive detection of the BP phase.
\bibitem{note3}  From the experimentally measureable column integrated density,
it is possible to reconstruct the full three-dimensional density
profile (M. W. Zwierlein, private discussion), particularly in the
case where the local density approximation holds well. The signal
of the BP phase becomes sharper in this full density profile.
\bibitem{note4}  The deep BEC region with $k_{F}a_{s}\ll 1$, however, is not a
good choice. The Cooper pairs become tightly bound there
($n_{\mathbf{k}}$ becomes very broad), and the relative change in
the momentum distribution becomes smaller. For $\left(
k_{F}a_{s}\right) ^{-1}=2$ or $3$, the size of the dip is
comparable with the width of the distribution $n_{\mathbf{k}}$,
and the Cooper pair wave function is significnatly modified by the
presence of the excess fermionic atoms.
\bibitem{note5}  Similar conclusion is drawn in Refs. \cite{7,8,10}.
A recent study using Monte Carlo simulation, however, suggests
that the BP phase might be stable at resonance \cite{mc}, but the
result is not yet conclusive due to the large error bar there.
\end{thebibliography}
\end{document}